\documentclass[11pt]{cernrep}
\usepackage{graphicx}

\newcommand{\Pt}{{E_t}}
\newcommand{\Et}{{E_t}}
\newcommand{\Ptg}{\Et^{\gamma}}
\newcommand{\ptg}{$\Et^{\gamma}$~}
\newcommand{\dphi}{\Delta\phi}
\newcommand{\phigj}{\phi_{(\gamma,jet)}}

\newcommand{\rrr}{\to}
\newcommand{\gpj}{~``$\gamma+jet$''~}

\newcommand{\lt}{\!<\!}
\newcommand{\gt}{\!>\!}

\begin{document}
\thispagestyle{empty}
 
\vskip-5mm
 
\begin{center}
{\Large JOINT INSTITUTE FOR NUCLEAR RESEARCH}
\end{center}
 
\vskip10mm
 
\begin{flushright}
JINR Communication \\
E2-2001-260 \\
hep-ex/0109001
\end{flushright}
 
\vspace*{3cm}
 
\begin{center}
\noindent
{\Large
{\bfseries Separation of quark and gluon jets in the direct photon production processes
at the LHC using the neural network approach.}}\\[10mm]
{\large D.V.~Bandourin$^{1}$, N.B.~Skachkov$^{2}$}
 
\vskip 0mm
 
{\small
{\it
E-mail: (1) dmv@cv.jinr.ru, (2) skachkov@cv.jinr.ru}}\\[3mm]
\large \it Laboratory of Nuclear Problems \\
\end{center}
 
\vskip 10mm
\begin{center}
\begin{minipage}{150mm}
\centerline{\bf Abstract}
~\\[1pt]
\noindent
A neural network technique is used to discriminate between quark and 
gluon jets produced in the $qg\to q+\gamma$ and $q\overline{q}\to g+\gamma$
processes at the LHC. 
Considering the network as a trigger and using the PYTHIA event generator and 
the full event fast simulation package for the CMS detector CMSJET 
we obtain signal-to-background ratios.
\end{minipage}
\end{center}       

\newpage
    
\setcounter{page}{1}    
 
\section{Introduction.}
%
There are  two QCD processes mainly contributing to the production of a direct photon:
the Compton-like process 
\\[-20pt]
\begin{eqnarray}
\hspace*{.74cm} qg\to q+\gamma 
\end{eqnarray}
\vspace{-4mm}
and the annihilation process\\[-10pt]
\begin{eqnarray}
\hspace*{.62cm} q\overline{q}\to g+\gamma. 
\end{eqnarray}                 
It was proposed in our paper \cite{SF}
to use the direct photon production processes to extract a gluon distribution function
in a proton $f^g(x,Q^2)$. It can be done by selecting those \gpj events which satisfy
the criteria pointed out in \cite{BKS1} and \cite{BKS5} to suppress the next-to-leading order 
diagrams with initial state radiation and the background to the direct photon production from the 
neutral decay channels of $\pi^0,\eta,K^0_s,\omega$ mesons and 
the photons radiated from a quark in the QCD 
processes with big cross sections (like $qg\to qg$, $qq\to qq$ and $q\bar{q}\to q\bar{q}$ 
scatterings).

A percentage of Compton-like process (1) (amounting to $100\%$ together with (2)) 
for different transverse energy $\Pt^{jet} (\approx \Ptg)$ and pseudorapidity 
$\eta^{jet}$ intervals are given in Table \ref{tab1}:\\[-20pt]
\begin{table}[h]
\begin{center}
\caption{ A percentage of the Compton-like process $q g\rrr \gamma +q$.}
\vskip.2cm
\begin{tabular}{||c||c|c|c|}                  \hline \hline
\label{tab1}
Calorimeter& \multicolumn{3}{c|}{$\Ptg$ interval ($GeV$)} \\\cline{2-4}
    part            & 40--50 & 100--120 & 200--240   \\\hline \hline
Barrel              & 89     &  84   &  78  \\\hline
Endcap+Forward      & 86     &  82   &  74  \\\hline 
\end{tabular}
\end{center}
\end{table}
~\\[-14pt]
In the table above the string ``Barrel'' corresponds to the Barrel region of the CMS calorimeter 
($|\eta|<1.4$) while  the string ``Endcap+Forward'' corresponds to the Endcap+Forward region 
($1.4<|\eta|<5.0$).

Thus, an admixture of the processes with a gluon jet in the final state grows from the left upper
corner to the right bottom one, i.e. with a jet energy.
Therefore, to collect a clean sample of $``\gamma+quark~ jet''$ events sample it is necessary
 to reject $``\gamma+gluon~ jet''$ events. This is most important in
the Endcap+Forward region for jets with $\Pt^{jet}>100~ GeV$ where the part of
the $``\gamma+gluon~ jet''$ events is more than $20\%$ and where one can reach the smallest
$x$ values of the gluon distribution function $f^g(x,Q^2)$ (see \cite{SF}). 

The idea of using the Artificial Neural Network (ANN) to discriminate quarks from gluons  
was widely discussed in the literature (\cite{JN} -- \cite{NN_IJ}). 
In \cite{GL_TR}, \cite{NN_IJ} the discrimination procedure is described
for $e^+e^-$ reactions at $\sqrt{s}=29,~92 ~GeV$ with three different Monte Carlo (MC)
generators: JETSET, ARIADNE and HERWIG. After testing with a middle point criterion the network
was able to classify correctly, on the average, $85\%$ of quark and gluon jets for a testing set. 
The MC 
independence of the results was also demonstrated by training with the MC data simulated by one 
generator and by testing with the MC data from another. 
We also refer to \cite{Kanda}, where MC independence (JETSET/HERWIG)
of the quark/gluon separation procedure based on the moment analysis of jet particles is presented.

In \cite{UseNN} the ANN was applied to a set of $p\bar{p}$ events at $\sqrt{s}=630 ~GeV$ 
generated with PYTHIA 
\cite{PYT}. The UA2 calorimeter geometry was used there to classify quark and gluon jets produced in 
the $qq\to qq,~q\bar{q}\to q\bar{q}$ and $gg\to gg$ QCD subprocesses alone. 
The $70-72\%$ classification ability with respect to the middle point criterion was reached there.
In this paper we use the ANN approach to get the most effective
 discrimination of quark and gluon jets in processes (1) and (2) selected by the cuts
given in Section 3 (and earlier in \cite{BKS1}, \cite{BKS5}). The close results were obtained
in \cite{GRAH} by using two- and three-layered network for quark/gluon jets classification
in the $p\bar{p}\to 2~ jets$ events at $\sqrt{s}=630 ~GeV$.

The study was carried out using the JETNET 3.0 package
\footnote{It is available via {\it anonymous} {\sf ftp} from {\sf thep.lu.se} or from 
{\sf freehep.scri.fsu.edu}.}
developed at CERN and the University of Lund \cite{JN}.

\section{Artificial Neural Network.}
%
\subsection{Generality and mathematical model of the neural network.}

ANNs are often used to optimize a classification (or pattern recognition) procedure
and was applied to many pattern recognition problems in high energy physics
(see \cite{UseNN} -- \cite{NN_IJ}, \cite{THES}, [17] -- [19]) 
with a notable success. They usually have more input than output nodes and thus may be viewed 
as performing dimensionality reduction of input data set.

The ANN approach is a technique which assigns objects to various classes. These objects can be
different data types, such as a signal and a background in our case. Each data type is assigned
to a class which in the context of the given paper is 0 for the background (gluon jet) and
1 for the signal (quark jet). Discrimination is achieved by looking at the class to which
the data belongs. The technique fully exploits the correlation among different variables
and provides a discriminating boundary between the signal and the background.

ANNs have an ability to learn, remember and create relationships amongst the data.
There are many different types of ANN but the feed forward types are most popular in
the high energy physics. Feed forward implies that information can only flow in one direction
and the output directly determines  the probability that an event characterized by some input
pattern vector $X(x_1,x_2,...x_n)$ is from the signal class.


The mathematical model of the Neural Network (NN) reflects three basic functions of a biological
neuron:\\[-6mm]
\begin{flushleft}
\parbox[r]{.53\linewidth}{
\begin{itemize}
\item sum up all the information arriving at inputs of the node/neuron;
\item if sum is greater than some threshold, fire neuron;
\item after firing, return to the initial state and send a signal to each of the
neighboring neurons in the network.
\end{itemize}
The neuron with these characteristics is known as an elementary perceptron.
The perceptron
is a simple feed forward system with several input connections and a single output connection.
}
\end{flushleft}
\begin{flushright}
  \begin{figure}[h]
\vskip-91mm
\hspace*{40mm} \includegraphics[width=0.99\textwidth,height=6.9cm]{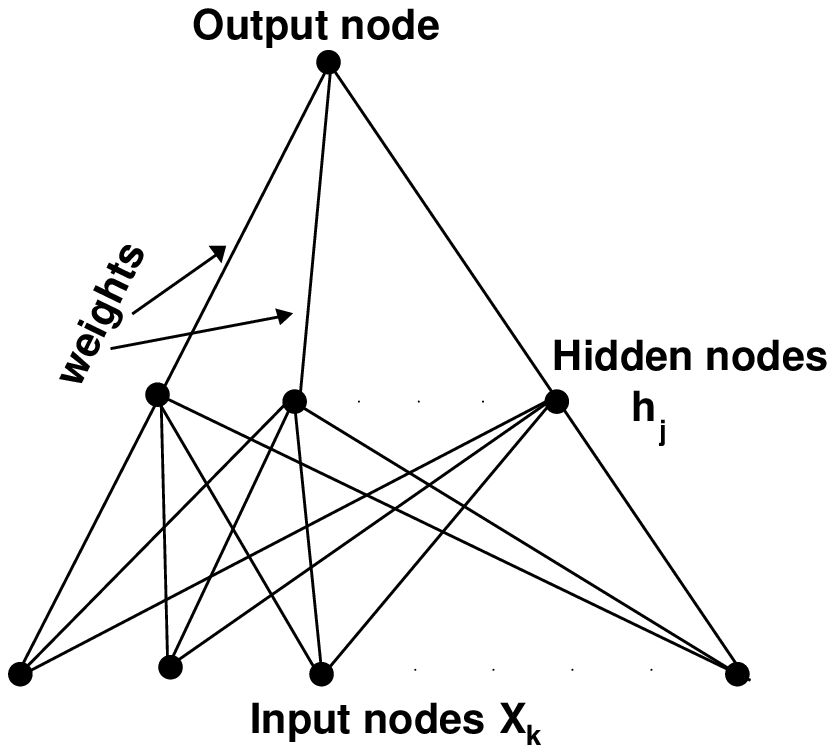}
\vskip-2mm
\hspace*{75mm} {\footnotesize Fig.~1. Neural network with one layer\\}
\hspace*{82mm} {\footnotesize of hidden units.}
    \label{fig:ANN}
  \end{figure}
\vskip-4mm
\end{flushright}
\setcounter{figure}{1}
\noindent
Mathematically the output can be written as\\[-20pt]
\begin{eqnarray}
O(x_1,x_2,...x_n) = g(\frac{1}{T}\sum\limits_{i} \omega_i x_i + \theta).
\label{eq:out1}
\end{eqnarray}
~\\[-15pt]
Here $g$ is a non-linear transfer function and typically takes the following form (sigmoid function)\\[-23pt]
\begin{eqnarray}
g=\frac{1}{1+e^{-2x}},
\label{eq:gx}
\end{eqnarray}
$(x_1,x_2,...x_n)$ is the input pattern vector, $O$ is the output, $\omega_i$ and $\theta$
are independent parameters called weights (which connect the input nodes to the output node)
and a threshold of the output node. $\beta=1/T$ is called inverse temperature
and defines the slope of $g$.

The pattern vector $x_i$ is multiplied by the connection weights $\omega_i$ so that each piece of
information appears at the perceptron as $\omega_i x_i$. Then the perceptron sums all the incoming
information to give $\sum \omega_i x_i$ and applies the transfer function $g$ to give the output
(see (\ref{eq:out1})).

In a feed forward NN a set of neurons has a layered structure. Figure~\ref{fig:ANN} shows
the feed forward the NN with one hidden layer that is used here.
In this case the output of NN is\\[-15pt]
\begin{eqnarray}
O(x_1,x_2,...x_n) = g(\frac{1}{T}\omega_j\sum\limits_{k} g(\frac{1}{T}
\sum\limits_{k} \omega_{jk} x_k + \theta_j) + \theta),
\label{eq:out2}
\end{eqnarray}
where $\omega_{jk}$ is the weight connecting the input node $k$ to the hidden node $j$ and
$\omega_j$'s connect the hidden nodes to the output node. $\theta_j$ and  $\theta$ are the thresholds
of the hidden and the output node respectively.

\subsection{Learning of the perceptron.}
%

The behavior of a perceptron is determined by independent parameters known as
weights and thresholds. The total number of independent parameters in a neural network
with a single layer is given by:
\begin{eqnarray}
N_{ind}=(N_{in}+N_{on})\cdot N_{hn} + N_{ht} + N_{ot}
\label{eq:n_ind}
\end{eqnarray}
where $N_{in}$ is a number of input nodes, $N_{on}$ is a number of output nodes,
$N_{hn}$ is a number of nodes in a hidden single layer, $N_{ht}$ is a number of thresholds in 
a hidden single layer, $N_{ot}$ is a number of output thresholds.

Learning is the process of adjusting these $N_{ind}$ parameters. During learning every 
perceptron is shown examples of what it must learn to interpret. It is fulfilled on the training set
consisting of two parts: training data (a collection of input patterns to the perceptron) and 
a training target, which is a desired output of each pattern.

Mathematically, the goal of training is to minimize a measure of the error. The mean squared
error function $E$ averaged over the training sample is defined by equation 
(\ref{eq:err})\\[-25pt]
\begin{eqnarray}
E=\frac{1}{2N_p} \sum\limits_{p=1}^{N_p}\sum\limits_{i=1}^{N} (O^{(p)}_i - t^{(p)}_i)^2,
\label{eq:err}
\end{eqnarray}
where $O_i$ is the output of the $i$th node of the NN in equation (\ref{eq:out2}); 
$t_i$ is the training target
(in our case, 0 for the background and 1 for the signal); $N_p$ is the number of patterns (events)
in the training sample; $N$ is the number of network outputs ($N=1$ for our case).

There are several algorithms for error minimization and weight updating. Most popular are 
{\bf Back propagation}, {\bf Langevin} and {\bf Manhattan} methods. In the last
one the weight is updated during the learning by the following rule 
\footnote{see \cite{CERN_NN} for a more complete description}:
~\\[-7mm]
\begin{eqnarray}
\omega_{t+1} = \omega_{t} + \Delta \omega \\[2pt]
\Delta \omega = -\eta \cdot sgn[\partial E/\partial \omega]
\end{eqnarray}
where $\omega$ is the vector of weights and thresholds used in the network;
$t~ (t+1)$ refers to the previous (current) training cycle and $\eta$ is the learning rate
which is decreased in the learning process.

\section{Event selection and Monte Carlo simulations for the ANN analysis.}
%
Our selection conditions for  \gpj events are based on the selection rules
chosen in \cite{BKS1} and \cite{BKS5}. We suppose the electromagnetic calorimeter (ECAL) size 
to be limited by $|\eta| \leq 2.61$ and the hadronic calorimeter (HCAL) is limited by 
$|\eta| \leq 5.0$ (the CMS geometry; see \cite{CMS_EC} and \cite{CMS_HC}),
where $\eta=-ln(tan (\theta/2))$ is a pseudorapidity 
%
\footnote{not to be confused with the learning rate also designated by $\eta$}
defined through
a polar angle $\theta$ counted from the beam line. In the plane
transverse to the beam line the azimuthal angle $\phi$ defines the
directions of $\vec{\Pt}^{jet}$ and $\vec{\Pt}^{\gamma}$.

\noindent
1. We select the events with one jet and one photon candidate with\\[-5pt]
\begin{equation}
\Pt^{\gamma} \geq 40 \; GeV~  \quad {\rm and} \quad \Pt^{jet} \geq 30 \;GeV.
\label{eq:sc1}
\end{equation}
A jet is defined here according to the PYTHIA jetfinding algorithm LUCELL
\footnote{PYTHIA's default jetfinding algorithm}.
The jet cone radius R in the $\eta-\phi$ space is
taken as $R=((\Delta\eta)^2+(\Delta\phi)^2)^{1/2}=0.7$.

\noindent
2. Only the events with ``isolated''
photons are taken to suppress the background processes. To do this, we

a) restrict the value of the scalar sum of $\Pt$ of hadrons and other particles surrounding
a photon within a cone of $R^{\gamma}_{isol}=( (\Delta\eta)^2 + (\Delta\phi)^2)^{1/2}=0.7$
(``absolute isolation cut")\\[-7pt]
\begin{equation}
\sum\limits_{i \in R} \Pt^i \equiv \Pt^{isol} \leq \Pt_{CUT}^{isol};
\label{eq:sc2}
\end{equation}
\vspace{-2.6mm}

b) restrict the value of a fraction (``relative isolation cut'')\\[-7pt]
\begin{equation}
\sum\limits_{i \in R} \Pt^i/\Pt^{\gamma} \equiv \epsilon^{\gamma} \leq 
\epsilon^{\gamma}_{CUT};
\label{eq:sc3}
\end{equation}

c) accept only the events having no charged tracks (particles) 
with $\Pt>1~GeV$ within the $R^{\gamma}_{isol}$ cone around a photon candidate.

\noindent
3. We consider the structure of every event with the photon
candidate at a more precise level of the 5$\times$5 crystal cells window (size of one CMS
HCAL tower) with a cell size of 0.0175$\times$0.0175. To suppress the background
events with the photons resulting from high-energy $\pi^0$, $\eta$, $\omega$
and $K_S^0$ mesons we require
\footnote{At the PYTHIA level of simulation this cut may effectively take into account 
the imposing of an upper cut on the HCAL signal in the tower behind
the ECAL $5\times 5$ crystal cells window hitted by the direct photon
(see \cite{GMO}).} that\\
{\it either} (a1) there is no high $\Pt$ hadron
in this 5$\times$5 crystal cells window
({\it at the PYTHIA level of simulation}):\\[-10pt]
\begin{equation} 
\Pt^{hadr} \leq 5~ GeV. 
\label{eq:sc5} 
\end{equation}
\hspace*{4.5mm} {\it or} (a2) the transverse energy deposited
in HCAL in the radius $R=0.7$ counted from the center of gravity of the HCAL tower just behind
the ECAL 5$\times$5 window, containing a direct photon signal, to be limited by
({\it at the level of the full event simulation}; see below) :\\[-10pt]
\begin{equation} 
\Et^{HCAL} \leq 1~ GeV. 
\label{eq:sc5} 
\end{equation}

\noindent   
4. The events with the vector $\vec{\Pt}^{jet}$ being ``back-to-back" to
the vector $\vec{\Pt}^{\gamma}$ within $\dphi$ in the plane transverse to the beam line 
with $\dphi$ defined by equation:\\[-5pt]
\begin{equation}
\phigj=180^\circ \pm \Delta\phi \quad (\Delta\phi =15^\circ, 10^\circ, 5^\circ)
\label{eq:sc7}
\end{equation}
($5^\circ$ is the size of one CMS HCAL tower in $\phi$)
for the following definition of the angle $\phigj$: 
$\vec{\Pt}^{\gamma}\vec{\Pt}^{jet}=\Pt^{\gamma}\Pt^{jet}\cdot cos(\phigj)$~
with ~$\Pt^{\gamma}=|\vec{\Pt}^{\gamma}|,~~\Pt^{jet}=|\vec{\Pt}^{jet}|$.
\noindent   

\noindent
5. To discard more the background events,
we choose only the events that do not have any other (except one jet)
minijet-like or cluster high $\Pt$ activity with the $\Pt^{clust}$
higher than some threshold $\Pt^{clust}_{CUT}$. Thus 
we select events with\\[-5pt]
\begin{equation}
\Pt^{clust} \leq \Pt^{clust}_{CUT},
\label{eq:sc8}
\end{equation}
where clusters are found by the same jetfinder LUCELL used to find a jet
in the same event.

The following values of cut parameters were used here:\\[-20pt]
\begin{eqnarray}
\Pt^{isol}_{CUT}=5\;GeV; \quad
{\epsilon}^{\gamma}_{CUT}=7\%; \quad
\dphi<15^{\circ}; \quad
\Pt^{clust}_{CUT}=10\;GeV.
\end{eqnarray}

To obtain the results of this paper we used two types of the generations: \\
(a) by PYTHIA alone,  based on 
the averaged calorimeter cell sizes $\Delta\eta\times\Delta\phi$:
$0.087\times0.087$ in the Barrel,
$0.134\times0.174$ in the Endcap and $0.167\times0.174$ in the Forward parts;\\
(b) by CMSJET -- the full-event fast Monte Carlo simulation package for a response in the CMS detector 
\cite{CMSJ} with the switched on calorimeter and magnetic field effects.

The following \ptg intervals were considered for both types of generations:
$40\!<\!\Ptg\!<\!50$, $100\!<\!\Ptg\!<\!120$ and $200\!<\!\Ptg\!<\!240 ~GeV$. 
Besides, for every \ptg interval we separate the
regions to which the jet belongs: Barrel ($|\eta^{jet}|\lt1.4$) and Endcap+Forward ($1.4\lt|\eta^{jet}|\lt4.5$). 
Since the jet is a spatially spread object, some energy leakage from one calorimeter
part to another is possible. To distinguish cases when a jet is in the Barrel or in
the Endcap+Forward regions the following restriction was added to cuts $1-5$:\\[-20pt]
\begin{eqnarray}
\Delta \Pt^{jet}/\Pt^{jet} = 0 - {\rm for~~ the~~ PYTHIA~~ level~~ study}; \\
\Delta \Pt^{jet}/\Pt^{jet} \leq 0.05 - {\rm for~~ the~~ CMSJET~~ level~~ study}. 
\end{eqnarray}
Here $\Delta \Pt^{jet}$ is the jet $\Pt$ leakage from that part of the calorimeter
in which the jet gravity center was found.

\section{Training and testing of ANN.}
%
There are two stages in the neural network analysis. The first is training of
the network and the second is testing. NN is trained with samples of signal and background
events and tested by using independent data sets. Training of the network corresponds to
step-by-step changing of the weights $\omega_{jk}$ such that a given input vector 
$X^{(p)}(x_1,x_2,...,x_n)$
produces an output value $O^{(p)}$ that equals the desired output or target value $t^{(p)}$
(see (5) and (7)).

The input parameters used in the 0th (input) layer of the network (Fig.~1)
were chosen as follows. In ``Set 01'' and ``Set 02'' we analyzed the jet information
obtained in PYTHIA simulation. In Set 01 we assigned $E_t$, $\eta$ and $\phi$ of the
first $E_t$ leading cell to the nodes $x_1$, $x_2$ and $x_3$ respectively. Then we
took the second leading cell and assign its $E_t$, $\eta$ and $\phi$ to the nodes
$x_4$, $x_5$ and $x_6$. The same was done for the remaining 13 cells. So, we
had 45 input nodes in total
\footnote{This input set is the same as in \cite{UseNN}. It was checked out that
variation down to 10 or up to 20 cells data at the input do not much affect the result.}.
In Set 02 we added 46th input node with a number of charged
tracks $N_{track}$ inside a jet with $\Pt^{ch}\gt1~GeV$. For ``Set 1'' and ``Set 2''
we repeated the previous procedure but with respect to the cells
 of jets found after the fast 
Monte Carlo simulation of the whole event by using CMSJET. Analogously, we had 45 and 46 
(+$N_{track}$ information) input nodes for Sets 1 and 2.

To ensure convergence and stability,
the total number of training patterns (events) must be significantly  
($20-30$ times) larger than the number of independent parameters (see (\ref{eq:n_ind})).
About 7000 signal (with a quark jet) and background (with a gluon jet) events
were chosen for the training stage, i.e. about 30 patterns per a weight.
\\[-20mm]
  \begin{figure}[htbp]
    \includegraphics[width=1.0\textwidth,clip]{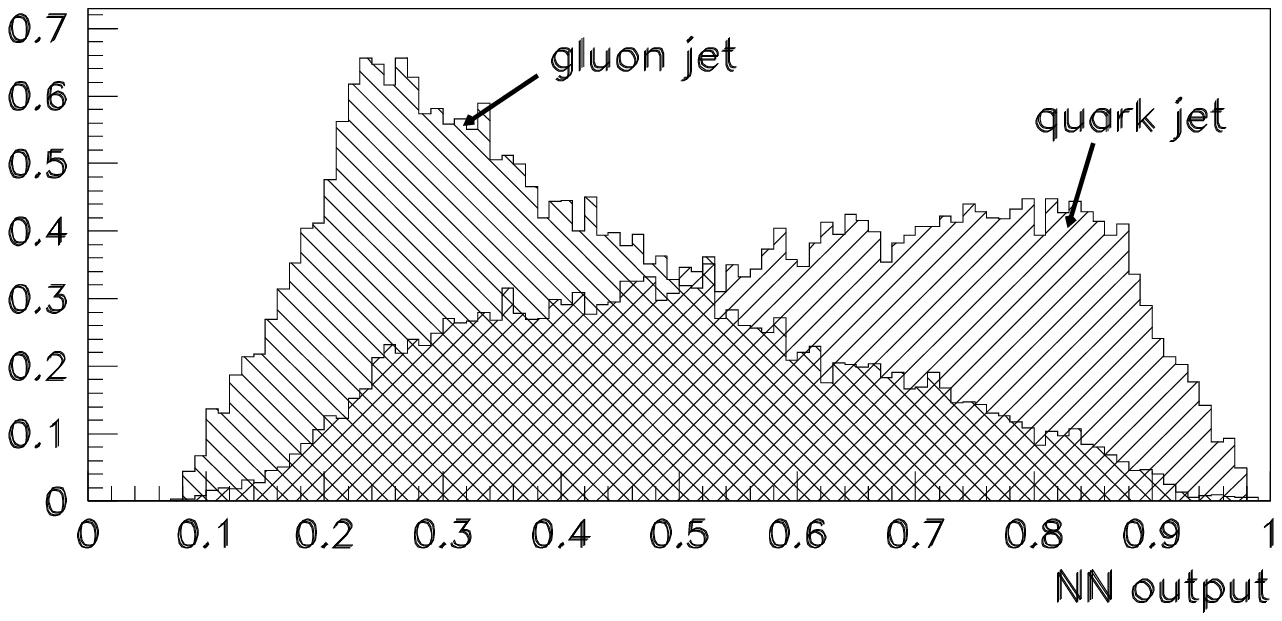}
    \vskip-6.5cm
    \caption{\footnotesize Neural network output for quark and gluon jets that were found in 
the Endcap+Forward region, $40\lt\Ptg\lt50 ~GeV$.}
    \label{fig:nn_out46}
  \end{figure}

After the NN was trained, a test procedure was implemented in which the events not
 used in the training were passed through the network. 
The same proportion of the signal and background events (about 7000 of each sort) was
used at the generalization stage. An output was provided for each event
and could be considered as a probability that an event is either from signal or background
sample. If the training is done correctly, the probability for an event being
 signal is high if the output $O$ is close to 1. And conversely, if the output $O$ is 
close to 0, it is more likely to be a background event (see Fig.~\ref{fig:nn_out46} 
for the case of jets found in the Endcap+Forward region and $40\!<\!\Ptg\!<\!50~GeV$ 
as an example of a typical NN output).

\section{The choice of neural network architecture and learning parameters.}
%

To investigate dependence of the separation possibility 
on the learning parameters, we trained a neural network
with 7000 signal events  and 7000 background events 
 found after the CMSJET simulation. In those events 
the direct photon $\Pt$ was chosen to be $100<\Pt^{\gamma}<120~GeV$ and jets were
found in the Barrel region. 

The network was tested with an independent set of 7000 signal
and background events. Sensitivity to different NN parameters was tested from the
point of view of the NN quark/gluon separation probability with respect to the ``0.5-criterion''
(point 

  \begin{figure}[htbp]
    \vskip-3cm
\hspace*{-5mm} \includegraphics[width=1.15\textwidth,height=6.5cm]{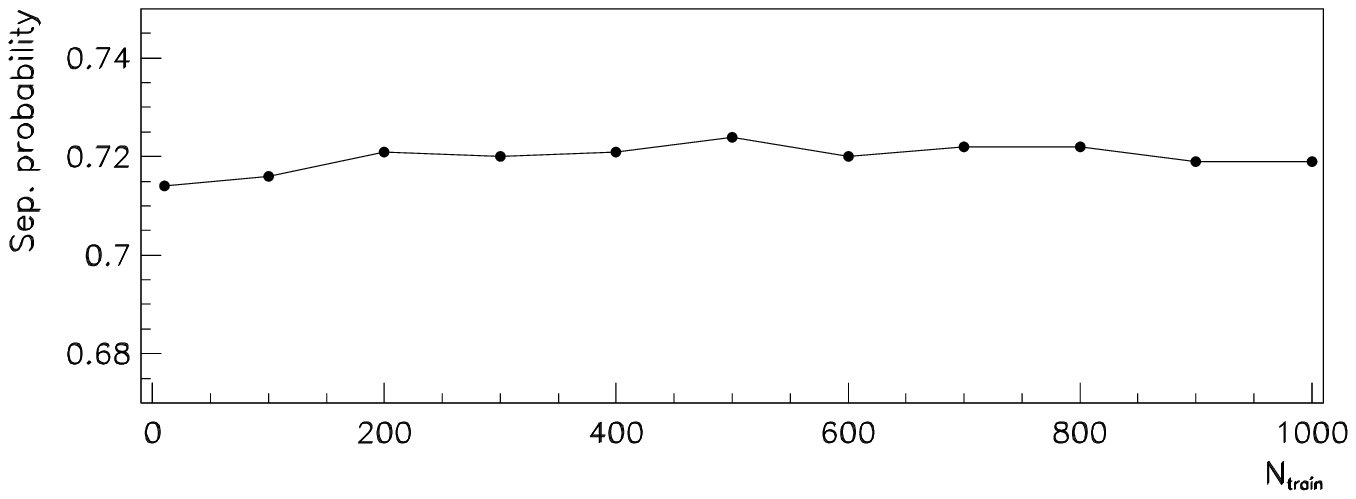}
    \vskip-0.9cm
    \caption{\footnotesize The quark/gluon separation probability using  ``0.5-criterion''
as a function of the number of training cycles $N_{train}$.}
    \label{fig:opt_t}
   \vskip-7mm
\hspace*{-7mm} \includegraphics[width=1.17\textwidth,height=14cm]{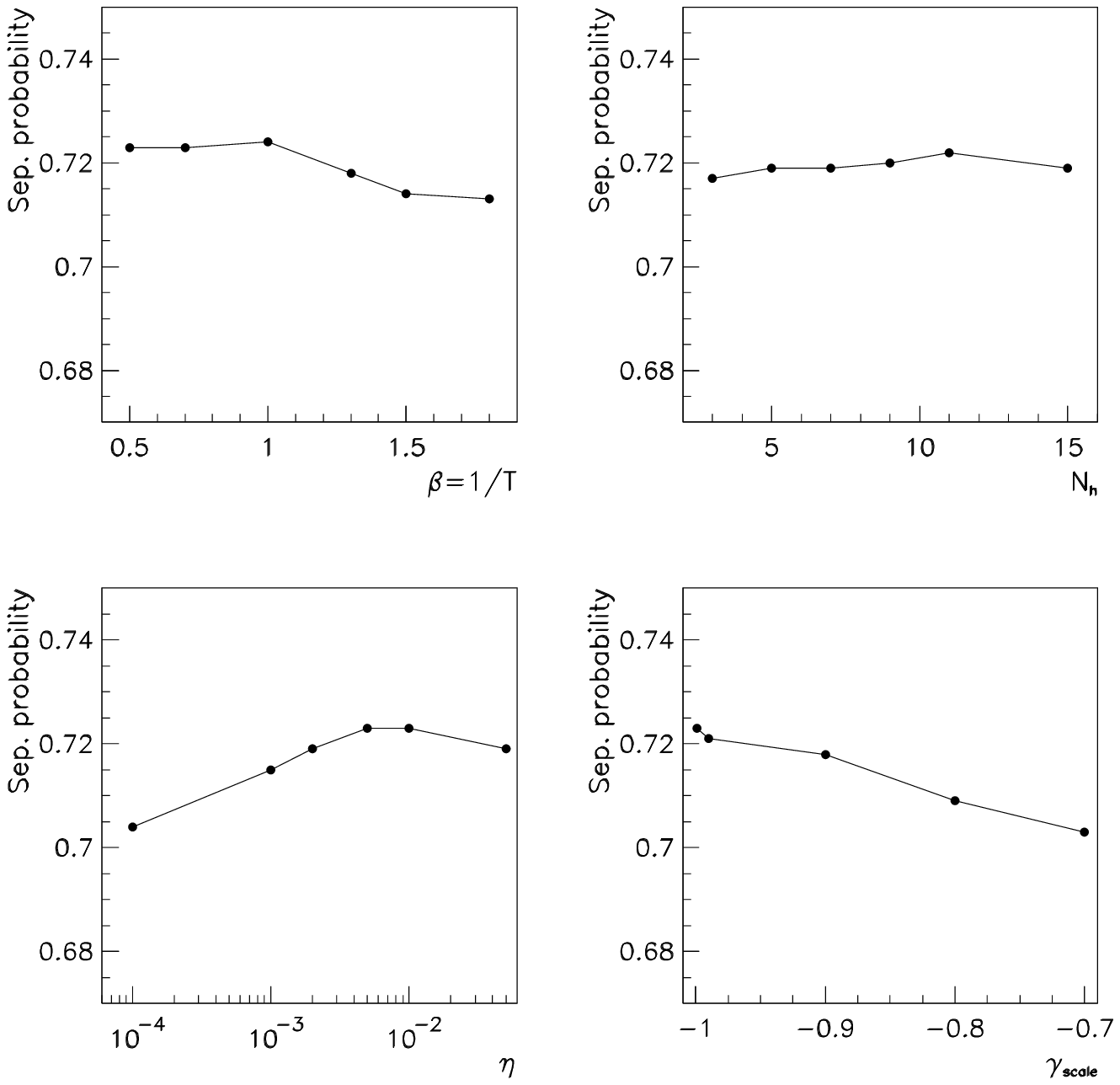}
    \vskip-.7cm
    \caption{\footnotesize The quark/gluon separation probability using the  ``0.5-criterion'' criterion
for various network parameters: inverse temperature $\beta=1/T$, the number of hidden nodes $N_{h}$,
learning rate $\eta$, $\eta$ updating scale parameter $\gamma_{scale}$. }
    \label{fig:opt_4x4}
  \end{figure}

\noindent
0.5 of the NN output). These parameters are listed below and 
the corresponding plots are given in Figs.~\ref{fig:opt_t} and \ref{fig:opt_4x4}.

\begin{itemize}
\item{{\it Number of training cycles}}\\
We varied the number of training cycles from 100 to 1000 to investigate the effect of
training on the network performance. The result shown in Fig.~\ref{fig:opt_t} 
indicates the network stability if more than 200 training cycles are used. 
\item{{\it Inverse temperature}}\\
The inverse temperature determines the steepness of the transfer function $g(x)$
(\ref{eq:gx}). On the left-hand upper plot of Fig.~\ref{fig:opt_4x4} 
the quark/gluon separation probability drops by $1\%$ as one goes from $\beta=0.5-1$ to
 $\beta=1.5-1.8$.

\item{{\it Number of hidden nodes}}\\
One hidden layer is  used here because it is sufficient for most classification
problems \cite{JN}. Sensitivity of the quark/gluon separation probability to a number of hidden
nodes $N_h$ was tested with $N_h=3-15$.  
All resulting points fall within $1\%$ ($71-72\%$) window (see Fig.~\ref{fig:opt_4x4})
\footnote{To be exact, a bit better result is achieved with $N_h=11$.}.

\item{{\it Learning rate $\eta$}}\\
The learning rate $\eta$ is a factor in updating the weights. We varied its value
between 0.0001 and 0.05 (see left-hand bottom plot in Fig.~\ref{fig:opt_4x4}).
 The value $\eta=0.005$  was chosen for our analysis.
\item{{\it Scale parameter $\gamma_{scale}$}}\\
The optimal learning rate $\eta$ varies during learning while the network converges
towards the solution. The scale factor for its changing is determined by the parameter
$\gamma_{scale}$. The right-hand bottom plot in Fig.~\ref{fig:opt_4x4} 
shows that the optimal performance is achieved at the default value $\gamma_{scale}=-1$.
\end{itemize}

~\\[-4mm]
\hspace*{9mm} As was mentioned above, the Manhattan updating method was used here during the training 
procedure. ~In~ Table \ref{tab6}~ for~ the case~ of jets~ in the Barrel region 
and $100\!<\!\Ptg\!<\!120 ~GeV$ (Set 2) this method is compared with
other updating algorithms with various  values of learning parameters:
learning rate $\eta$ (Backpropagation, Langevin) and noise term $\sigma$
(Langevin).  It is seen that by varying $\eta$ and $\sigma$ from their
~\\[-13mm]
\begin{table}[h]
\begin{center}
\vskip5mm
\caption{\footnotesize A dependence of the separation probability ($\%$) using
 ``0.5-criterion'' on the method. CMSJET, Set 2, Barrel region, $100<\Pt^{\gamma}<120~GeV$.}
\vskip-1mm
\begin{tabular}{||c||c|c|c||c|c|c||}                  \hline \hline
\label{tab6}
Method  &\multicolumn{3}{c||}{Backpropagation}&\multicolumn{3}{c||}{Langevin}
\\\cline{1-7}
Parameters &$\eta\!=\!1.$&$\eta\!=\!0.5$& $\eta\!=\!0.1$-
&$\eta\!=\!1.0$& $\eta\!=\!0.1\!-\!0.01$&$\eta\!=\!0.01$    \\
& & &$\eta\!=\!0.001$ 
&$\sigma\!=\!0.01$&$\sigma\!=\!0.01$&$\sigma\!=\!0.001$ \\\hline \hline
Probab.($\%$)& 51 & 68 & 71 & 69 & 70 & 71  \\\hline
\end{tabular}
\end{center}
\end{table}

\noindent
default values in the JETNET package (the first column for each algorithm)
one can approximately reach the value of the separation probability
obtained by using the Manhattan algorithm ($72\%$).

\section{Description of the results.}
%

As an example of the ``0.5-criterion'' application, Table~\ref{tab2} presents
the discrimination powers obtained after the simulation at the PYTHIA level
and events selection according to the cuts $(10)-(18)$ of Section 3
for three various intervals of the direct photon $\Pt$.
\\[-25pt]
\begin{table}[h]
\begin{center}
\caption{\footnotesize The quark/gluon separation probability ($\%$) using ``0.5-criterion''.
Barrel and Endcap+Forward regions. PYTHIA level simulation.}
\vskip.1cm
\begin{tabular}{||c|c|c|c|c||}                  \hline \hline
\label{tab2}
Simulation& Set  & \multicolumn{3}{c|}{$\Pt^{\gamma}$ interval ($GeV)$} \\\cline{3-5}
 type  & No.  & 40 -- 50 & 100 -- 120 & 200 -- 240  \\\hline \hline
Barrel & 01   & 74     &  76   &  79 \\\cline{2-5}
       & 02   & 75     &  77   &  82 \\\hline
Endcap+& 01   & 70     &  69   &  69\\\cline{2-5}
Forward& 02   & 73     &  74   &  75\\\hline 
\end{tabular}
\end{center}
\vskip-3mm
\hspace*{20mm}\footnotesize{The error is of order of $1.5-2\%$ for all numbers in the table above.}
\vskip-2mm
\end{table}

We see that by using the ``0.5-criterion'' for the ANN output, when the output node value $O\!>\!0.5$ is
interpreted as a quark jet and $O\!<\!0.5$ as a gluon jet, the network correctly classifies 
$75-82\%$ ($73-75\%$) of jets at the PYTHIA level in the Barrel (Endcap+Forward) region with 
the input data that correspond to Set 01 and Set 02 (see Section 4). 
The separation probability is seen to grow by  $1-3\%$ 
after introducing the information on the number of tracks $N_{track}$ in the Barrel region. 
The analogous increase for the Endcap+Forward region is $3-6\%$. 

To give an understanding of such an improvement we plot, as an example, a distribution of 
the number of events over the number of tracks with $\Pt\gt1~GeV$ in quark and gluon jets, 
i.e. $N_{track}^q$ and $N_{track}^g$, for $40\!<\!\Pt^{\gamma}\!<\!50 ~GeV$ and 
$200\!<\!\Pt^{\gamma}\!<\!240 ~GeV$ in the Endcap+Forward region (Fig.~\ref{fig:ntr})
\footnote{For a comparison, see also quark and gluon jet multiplicities found
in experiments at DELPHI \cite{DELPHI}, OPAL \cite{OPAL} and D0 \cite{D0} collaborations.}.
Due to the larger probability of bremsstrahlung from a gluon than from a quark we obtain
the $\left<N_{track}^g\right>/\left<N_{track}^q\right>$
ratio equal to 1.27 for $40<\Pt^{\gamma}<50 ~GeV$ and 1.46 for $200<\Pt^{\gamma}<240 ~GeV$.
~\\[-5mm]
\begin{center}
  \begin{figure}[htbp]
   \vskip-1.5cm
\hspace*{-5mm} \includegraphics[width=1.17\textwidth,height=6.8cm]{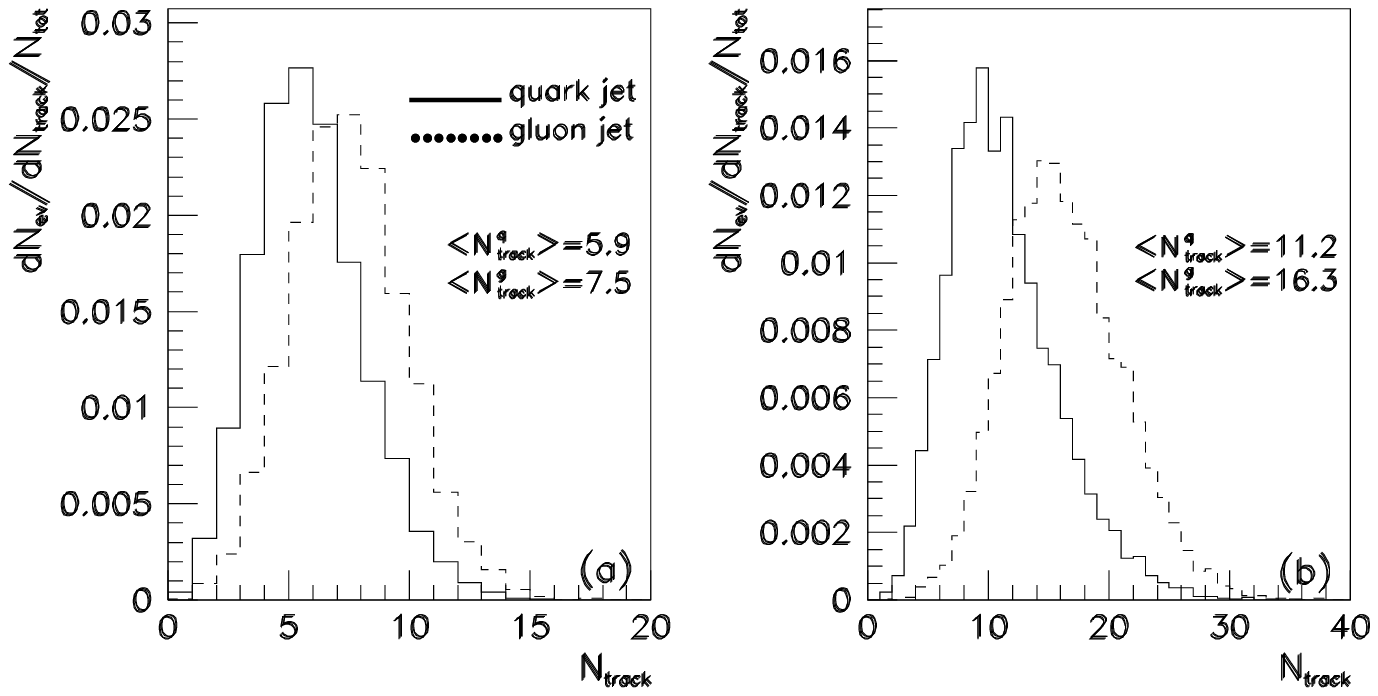} 
    \vskip-1.1cm
    \caption{\footnotesize Distribution over the number of charged tracks with 
$\Pt^{ch}\gt1~GeV$ for jets found in the Endcap+Forward region: $40\lt\Ptg\lt50 ~GeV$~ (a)  
and $200\lt\Ptg\lt240 ~GeV$~ (b).}
    \label{fig:ntr}
  \end{figure}
 \vskip-12mm
\end{center}

\begin{figure}[htbp]
    \includegraphics[width=1.0\textwidth,height=17cm]{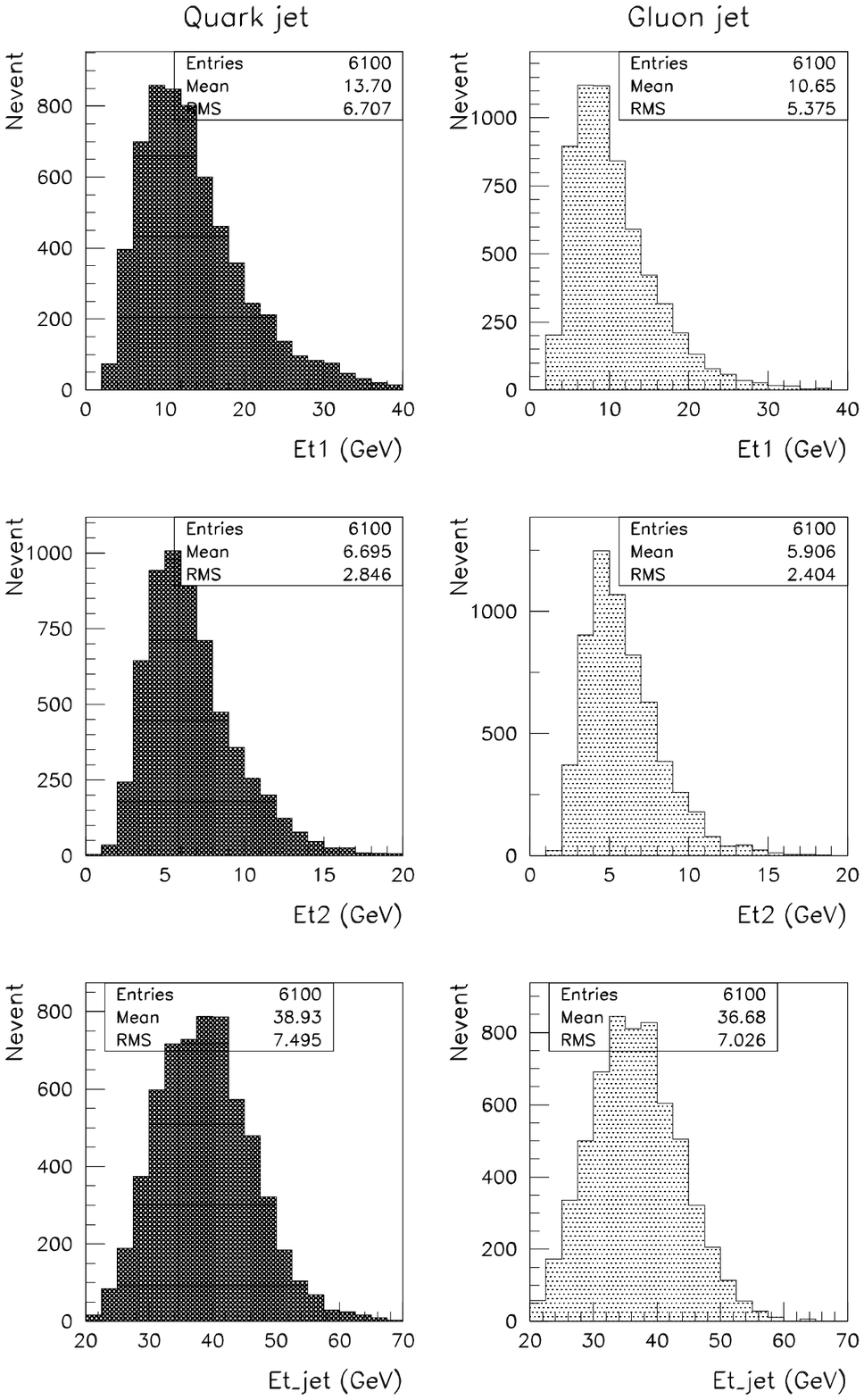}
    \vskip-.7cm
    \caption{\footnotesize Distribution over $Et$ of leading cell ($Et1$), $Et$ of next-to-leading cell
($Et2$) and $Et$ of the full quark and gluon jets. CMSJET, Endcap+Forward, $40\lt\Ptg\lt50 ~GeV$.}
    \label{fig:nc40}
  \end{figure}

\begin{figure}
    \includegraphics[width=1.0\textwidth,height=17cm]{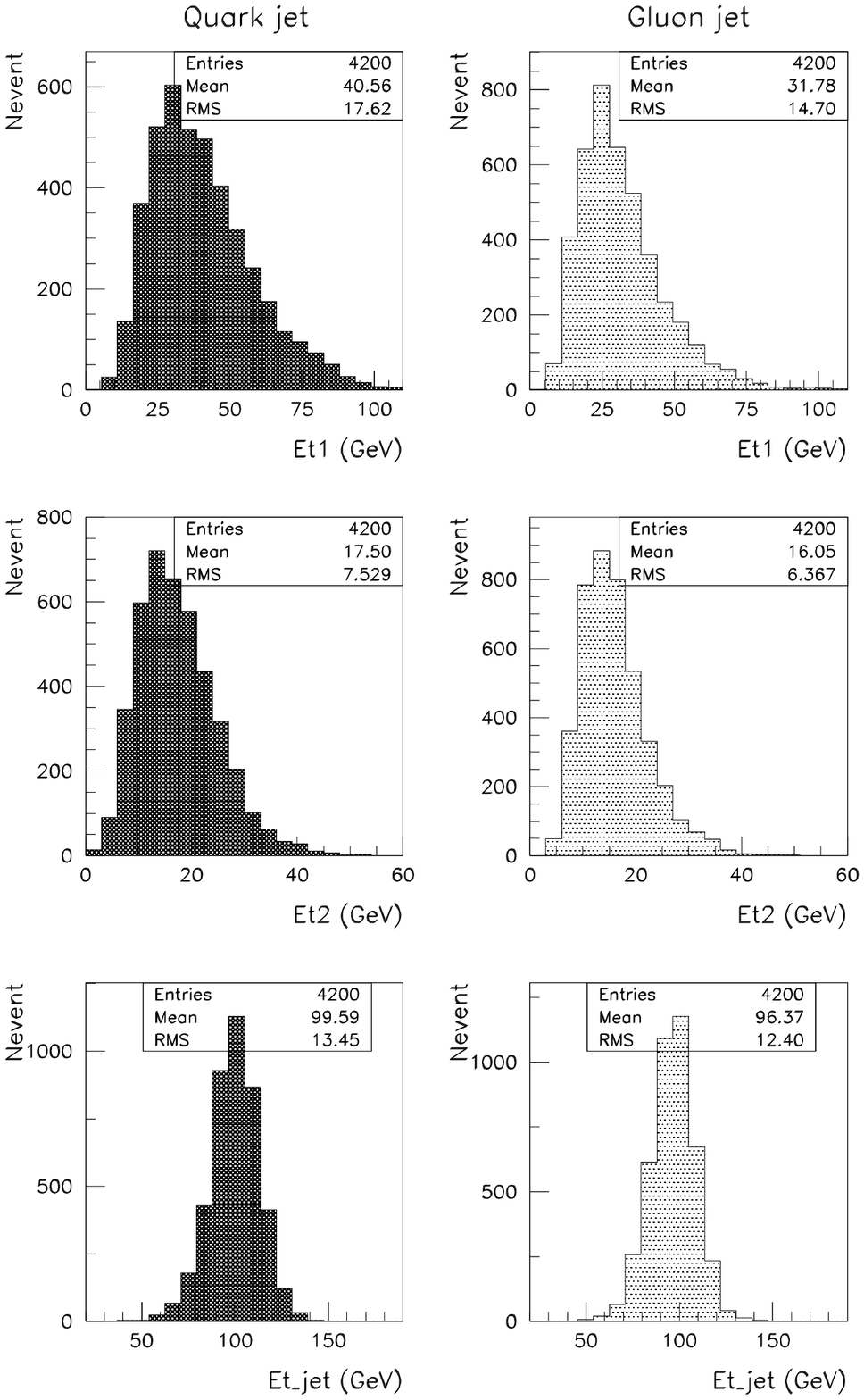}
    \vskip-.7cm
    \caption{\footnotesize Distribution over $Et$ of leading cell ($Et1$), $Et$ of next-to-leading cell
($Et2$) and $Et$ of the full quark and gluon jets. CMSJET, Endcap+Forward, $100\lt\Ptg\lt120 ~GeV$.}
    \label{fig:nc100}
  \end{figure}

\begin{figure}
    \vskip-1.3cm
   \hspace*{-9mm} \includegraphics[width=1.2\textwidth]{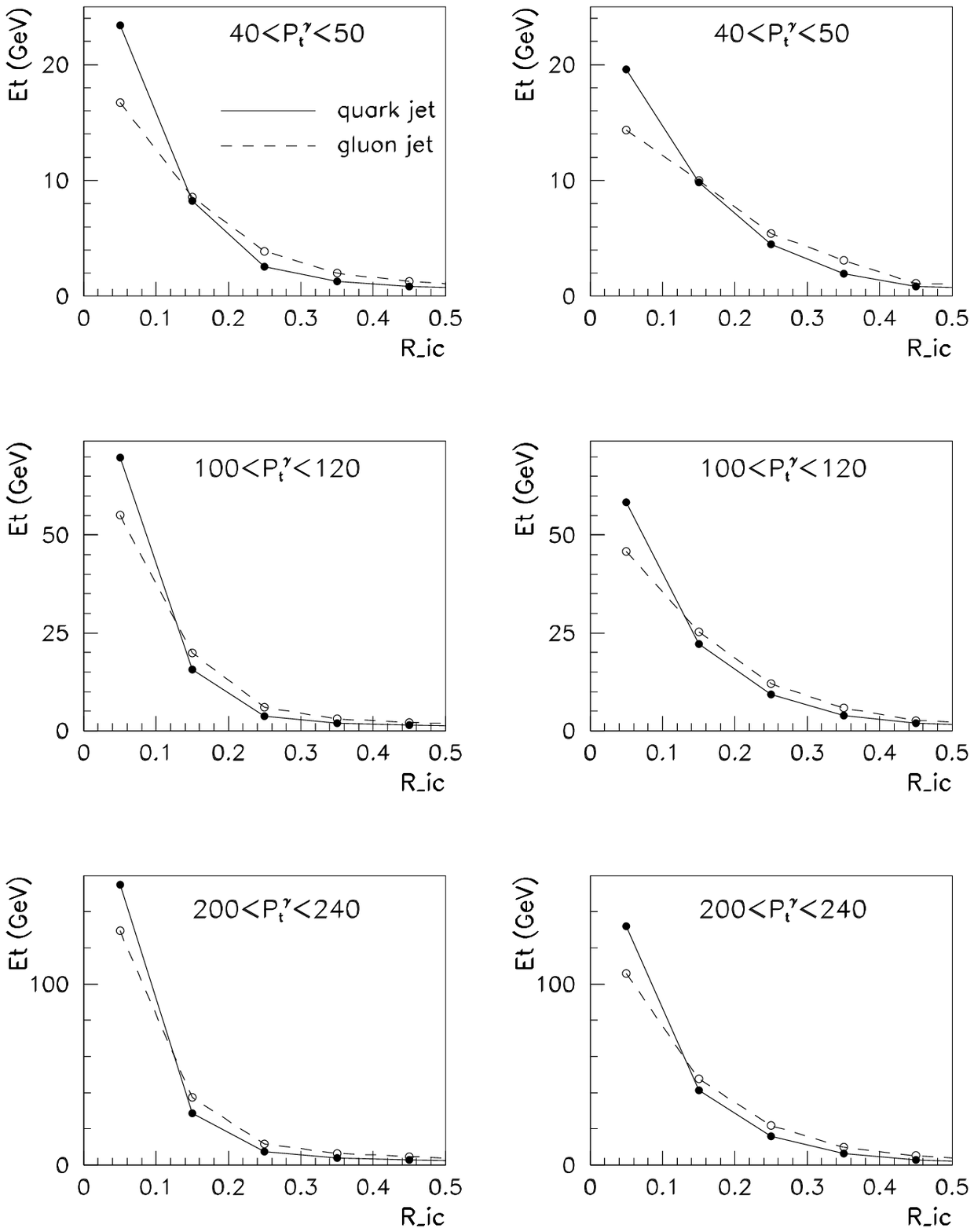}
    \vskip-.7cm
    \caption{\footnotesize Distribution of $E_t$ over the distance $R_-ic$ (in the $\eta-\phi$ space)
from the initiator cell inside quark (solid line) and gluon (dashed line) jets. The left-hand column
corresponds to the Barrel region and the right-hand to the Endcap+Forward region. The CMSJET simulation.}
    \label{fig:etr_be}
  \end{figure}

Figures~\ref{fig:nc40} -- \ref{fig:etr_be} obtained after the full event simulation with the help of
CMSJET also explain the choice of the variables at the input
 to NN in Section 4. As is seen from Figs.~\ref{fig:nc40} and \ref{fig:nc100},
$\Et$ of the leading cell (``$Et1$'' in the plots) in a quark jet is, on the average, 
 $25-30\%$ greater than in a gluon jet. The difference in $\Et$ for the next-to-leading cells 
(``$Et2$'' on the plots) in quark and gluon jets is  about $10-20\%$ 
(it is smaller for jets with a higher $\Et$). $\Et$ of a complete quark jet is 
also greater than $\Et$ of a complete gluon jet (by $4-10\%$). Again,
the difference becomes smaller with growing jet $\Et$.

Figure~\ref{fig:etr_be} shows a distribution of the averaged $\Et$ in quark and
gluon jets over the distance from the jet $\Et$ leading cell $R_-ic$ for all $\Ptg$
intervals and calorimeter regions considered in this paper. One can note
that in all cases the averaged $\Et$ in a quark jet up to $R_-ic\approx 0.12-0.14$
is greater than in a gluon jet and, vice versa, the averaged $E_t$ in a quark jet
for $R_-ic\geq 0.14$ is lower than in a gluon jet. 

It is more useful for practical applications to investigate the Signal/Background ratios 
for different the NN output thresholds
\footnote{not only for the point 0.5 as in Table 3 above}.
This analysis was done after the full simulation with CMSJET and event selection according to
cuts $(10)-(19)$. 

The Signal/Background ratios corresponding to the ``Set 2'' input NN 
information are given in Table \ref{tab4} for three $\Ptg$ intervals and two calorimeter regions. 
As a complement to Table \ref{tab4},
in Fig.~\ref{fig:sepp_eff} shows the quark selection and gluon rejection efficiencies in the case
of the full simulation for the same $\Ptg$ intervals and calorimeter regions.
 \\[-22pt]
\begin{table}[h]
\begin{center}
\caption{\footnotesize Signal/Background. The full event simulation using CMSJET. Set 2.}
\vskip.1cm
\begin{tabular}{||c||c|c|c|c|c|c|c|}                  \hline \hline
\label{tab4}
$\Ptg$&  &\multicolumn{6}{c|}{NN output cut} \\\cline{3-8}
$(GeV)$&Region & 0.3 & 0.4 & 0.5 & 0.6 & 0.7 & 0.8  \\\hline \hline
40 -- 50   & Barrel    & 1.45  & 1.91 & 2.40 & 3.11 & 4.19 & 6.16  \\\cline{2-8}
         & Endcap+Forward & 1.41  & 1.81 & 2.38 & 3.10 & 4.04 & 5.85 \\\hline
100 -- 120   & Barrel    & 1.72  & 2.63 & 3.26 & 4.04 & 4.59 & 6.37\\\cline{2-8}
         & Endcap+Forward & 1.75  & 2.19 & 2.95 & 3.61 & 4.21 & 5.41 \\\hline
200 -- 240  & Barrel    & 1.76  & 2.37 & 3.35 & 4.26 & 5.56 & 7.36 \\\cline{2-8}
         & Endcap+Forward & 1.64  & 2.40 & 3.17 & 4.16 & 5.39 & 7.45 \\\hline
\end{tabular}
\end{center}
\vskip-5mm
\end{table}

The Signal/Background ratio grows both with growing NN output threshold value and with 
increasing $\Ptg$ value (see Table \ref{tab4}). So, it grows from 2.4 to 3.2 at the NN output 
cut $O\!>\!0.5$ and from 4.0 to 5.4 at $O\!>\!0.7$ for the Endcap+Forward region. 
The curves in Fig.~\ref{fig:sepp_eff} show that for the last cut ($O\!>\!0.7$) about $38\%$ and $44\%$
of the events with the quark jet are selected for $40\!<\!\Ptg\!<\!50 ~GeV$ and $200\!<\!\Ptg\!<\!240 ~GeV$, 
respectively, while about $66-67\%$ of the events with quark jet are selected at $O>0.5$ for the both 
\ptg intervals and the same calorimeter region.

The Signal/Background ratio dependence on the NN output cut at the PYTHIA level is presented in 
Figs. \ref{fig:sb_hb} and \ref{fig:sb_hef}.

It is also important for practical realizations to know a dependence of the Signal/Background ratios 
on the quark jet selection efficiencies.
This dependence is plotted in Fig.~\ref{fig:qeff_sb0} for two extreme considered in this paper 
intervals $\Ptg$  and two 
calorimeter regions. We present two curves obtained with Set 1 and Set 2 of input information after the
full CMSJET event simulation (thin and thick solid lines) and one curve (dotted line)
obtained with Set 02 after event simulation at the PYTHIA level.


%
\section{Some additional remarks.}
%
The results obtained with the quark and gluon jets found in the CMSJET simulation were compared
with the results obtained after passing the quark  and gluon  jet particles through the electromagnetic
(ECAL) and hadronic (HCAL) calorimeters in the CMSIM package \cite{CMSIM}. The discrimination 
probabilities obtained after the cell analysis in CMSIM are found to be in good agreement
(up to $1-2\%$) with those obtained in CMSJET. It was also found  
that almost the same discrimination powers can be achieved both in CMSET and in CMSIM
 by using the network input information about $E_t$ of the first, $E_t$-ordered 
15 ECAL and 15 HCAL cells (i.e.~30 input nodes) instead of 45 input nodes as considered above
(see Section 4).
~\\[-6mm]
 \begin{center}
  \begin{figure}[htbp]
   \vskip-1.5cm
\hspace*{-5mm} \includegraphics[width=1.15\textwidth, height=11cm]{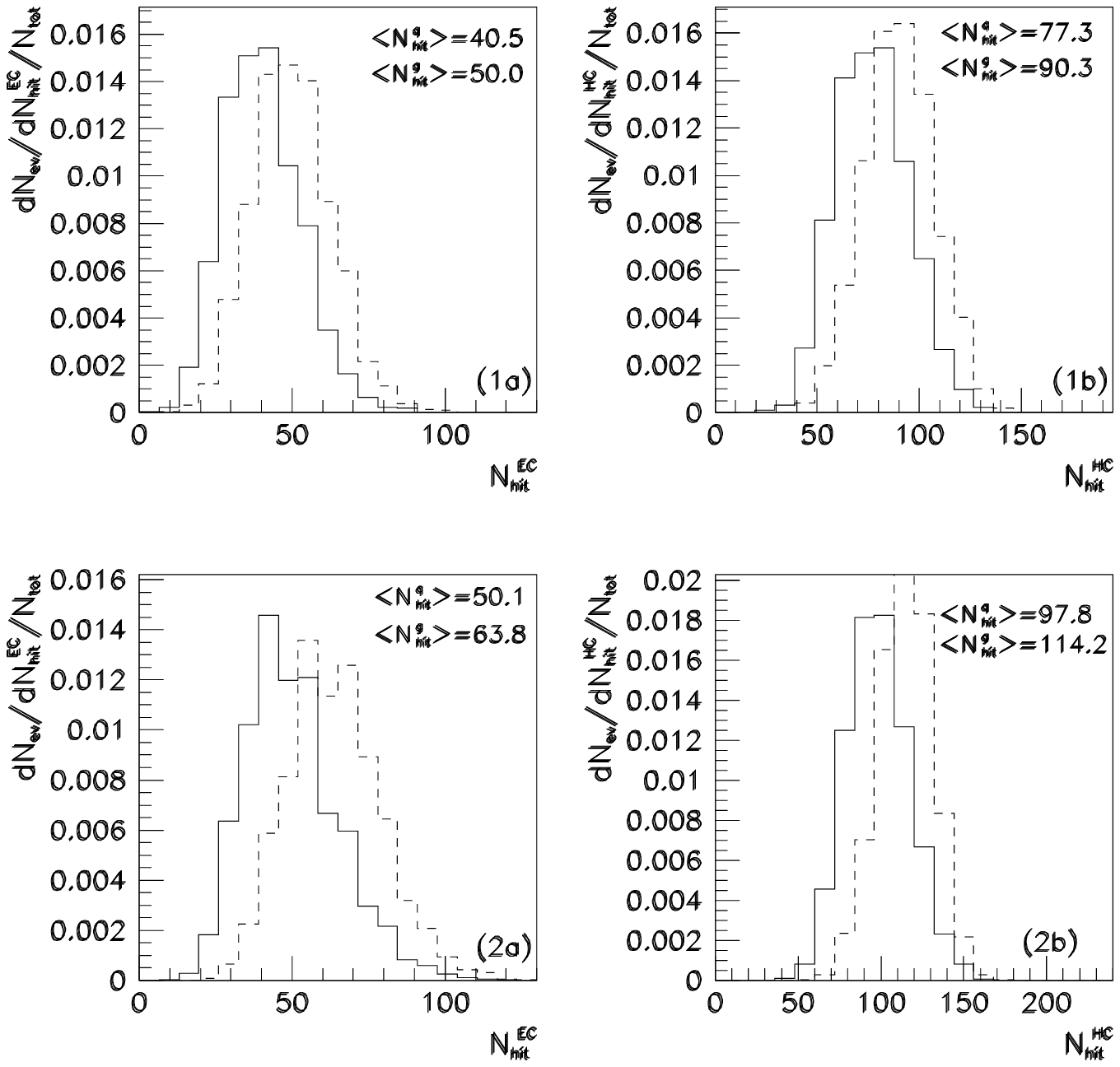}
    \vskip-0.9cm
    \caption{\footnotesize Distribution over the number of ECAL (plots 1a and 2a) and HCAL jet
cells (plots 1b and 2b), $N^{EC}_{hit}$ and $N^{HC}_{hit}$, 
for jets found in the Barrel region: $40\lt\Ptg\lt50 ~GeV$~ (1a, 1b)  
and $200\lt\Ptg\lt240 ~GeV$~ (1b, 2b).}
    \label{fig:nhit}
\vskip-12mm
  \end{figure}
\end{center}

The sensitivity of the network to some parameters is also noteworthy.
So, the network is able to classify correctly quark  and gluon jets with respect to the ``0.5-criterion''
in $65\%$ ($67\%$) of events with $40\!<\!\Ptg\!<\!50 ~GeV$ ($200\!<\!\Ptg\!<\!240 ~GeV$)
by using the $N_{track}$ variable alone. These results can be improved by $2-3\%$ if we also add
to $N_{track}$ two more input variables: the numbers of activated cells (towers) in the ECAL and the HCAL
belonging to quark and gluon jets. The distributions over those numbers are shown in
Fig.~\ref{fig:nhit} for two $\Ptg$ intervals: $40\!<\!\Ptg\!<\!50 ~GeV$ and 
$200\!<\!\Ptg\!<\!240 ~GeV$. We see that mean number of the activated cells in 
the ECAL for the case of gluon jets 
$\left<N_{hit}^g\right>$ exceeds that for the case of quark jets $\left<N_{hit}^q\right>$ 
by a factor of 1.23 for $40\!<\!\Ptg\!<\!50 ~GeV$.
This difference grows up to the factor of 1.28 for $200\!<\!\Ptg\!<\!240 ~GeV$.
And in both intervals the ratio of the mean numbers of the activated cells
$\left<N_{hit}^g\right>$/$\left<N_{hit}^q\right>$ in the HCAL is about $1.16-1.17$.

\noindent
{\bf Acknowledgements.} \\[3pt]                                             
We are greatly thankful to George Gogiberidze (JINR, Dubna) for the helpful discussions on 
some ideas and methods of the artificial neural network usage for pattern recognition tasks.

  \begin{figure}
   \vskip-15mm
\hspace*{-4mm} \includegraphics[width=1.15\textwidth,height=18cm]{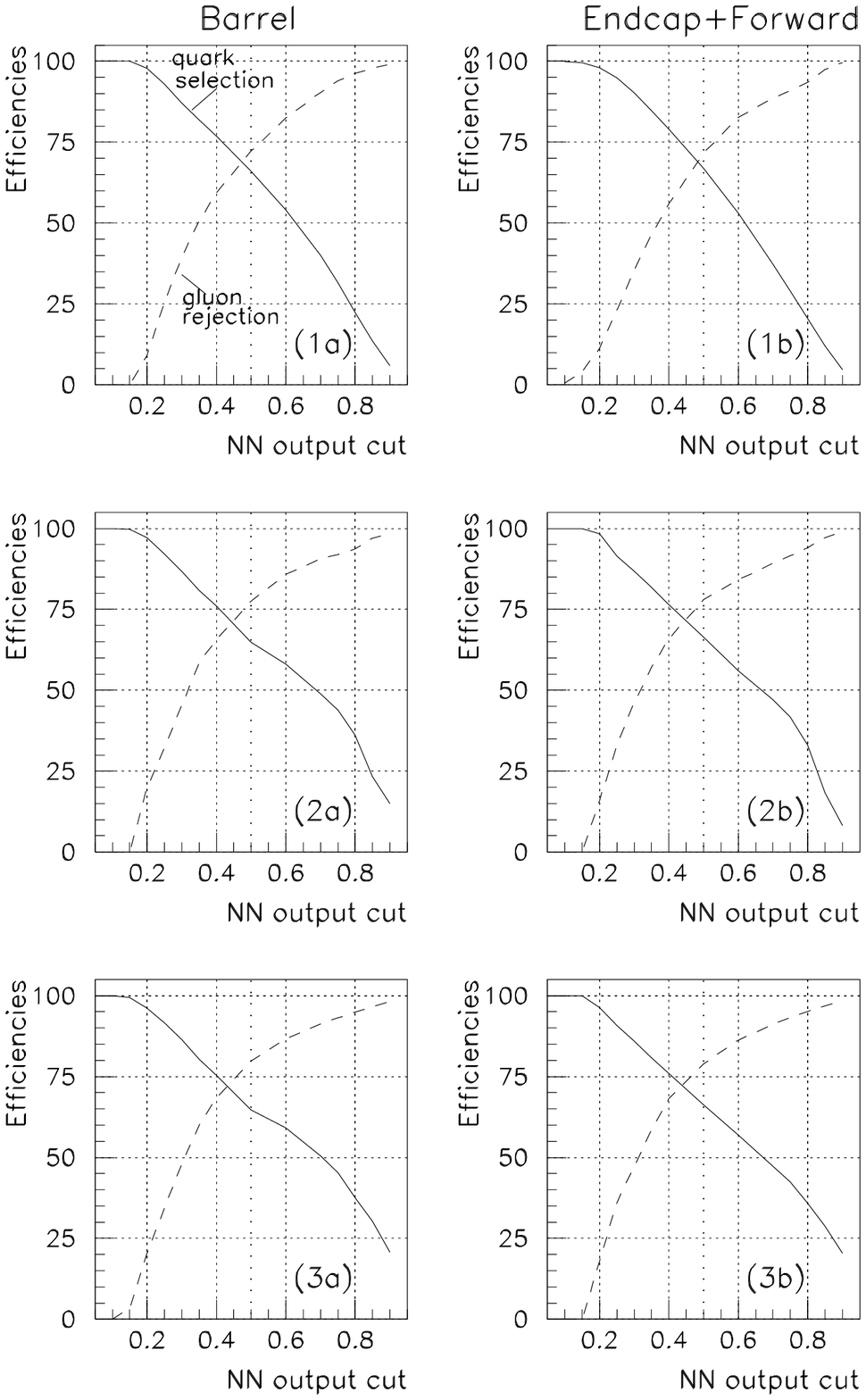}
    \vskip-5mm
    \caption{\footnotesize  Quark jet selection and gluon jet rejection efficiencies as a function  of neural network output cut. Left-hand  (1a, 2a, 3a) and right-hand 
(1b, 2b, 3b) columns correspond to the Barrel and the Endcap+Forward regions respectively. 
The first row plots (1a, 1b) are distributions for events selected with $40\!<\!\Pt^{\gamma}\!<\!50 ~GeV$, 
in the second (2a, 2b) with $100\!<\!\Pt^{\gamma}\!<\!120 ~GeV$ and in the third (3a, 3b) 
with $200<\Pt^{\gamma}<240 ~GeV$. Set 2.}
    \label{fig:sepp_eff}
  \end{figure}
  \begin{figure}
    \vskip-23mm
\hspace*{-5mm} \includegraphics[width=1.15\textwidth,height=18cm]{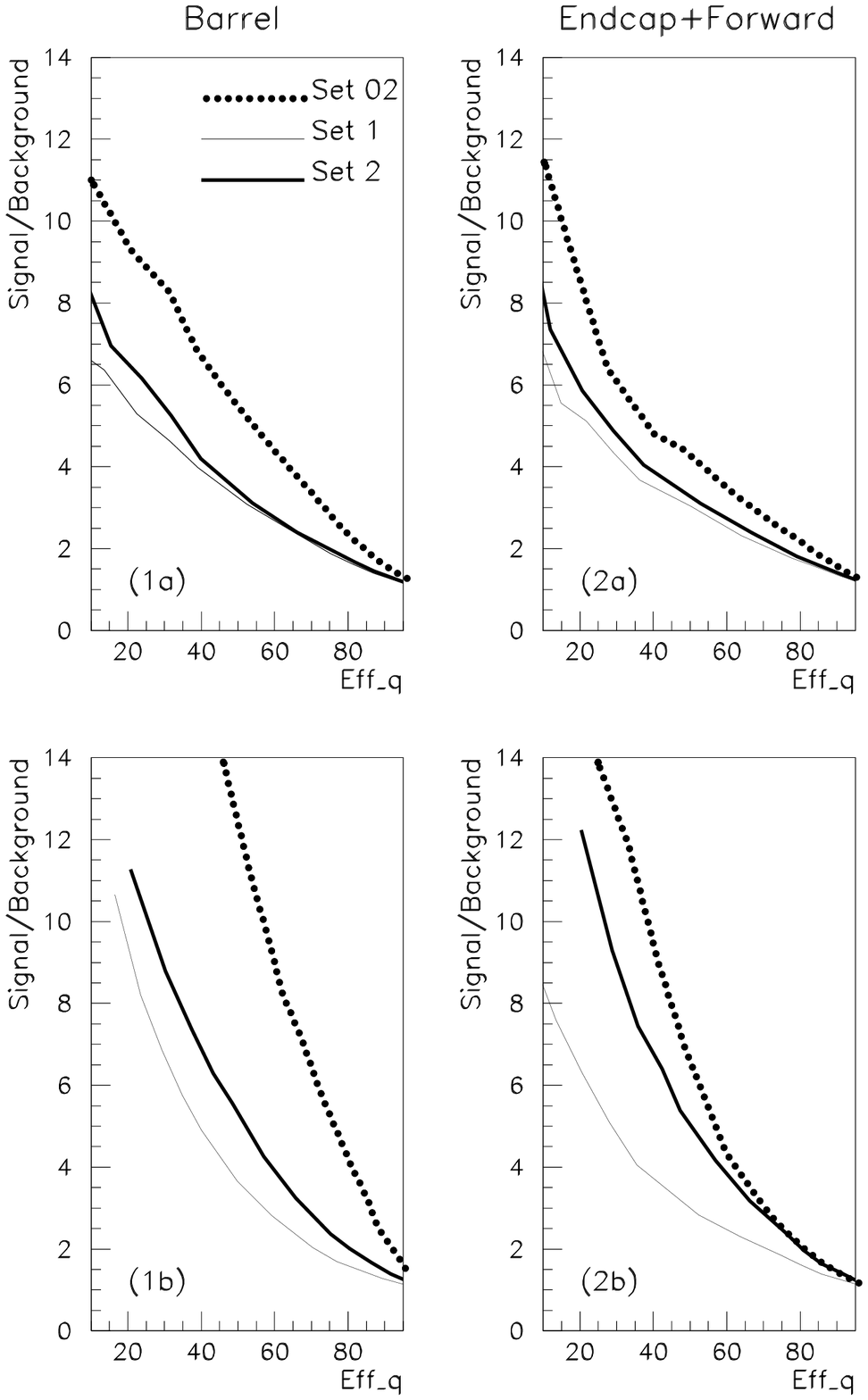}
    \vskip-7mm
    \caption{\footnotesize Signal to background ratio via quark jet selection efficiency.
The left-hand column (1a, 1b) correspond to the events with jets found in the Barrel and
the right-hand (2a, 2b) correspond to the events with jets found in the Endcap+Forward region. 
In the first row (1a, 2a) are distributions for events selected with $40\!<\!\Ptg\!<\!50 ~GeV$ 
and in the second with $200\!<\!\Ptg\!<\!240 ~GeV$.}
    \label{fig:qeff_sb0}
  \end{figure}
  \begin{figure}
    \includegraphics[width=1.0\textwidth,height=17.5cm]{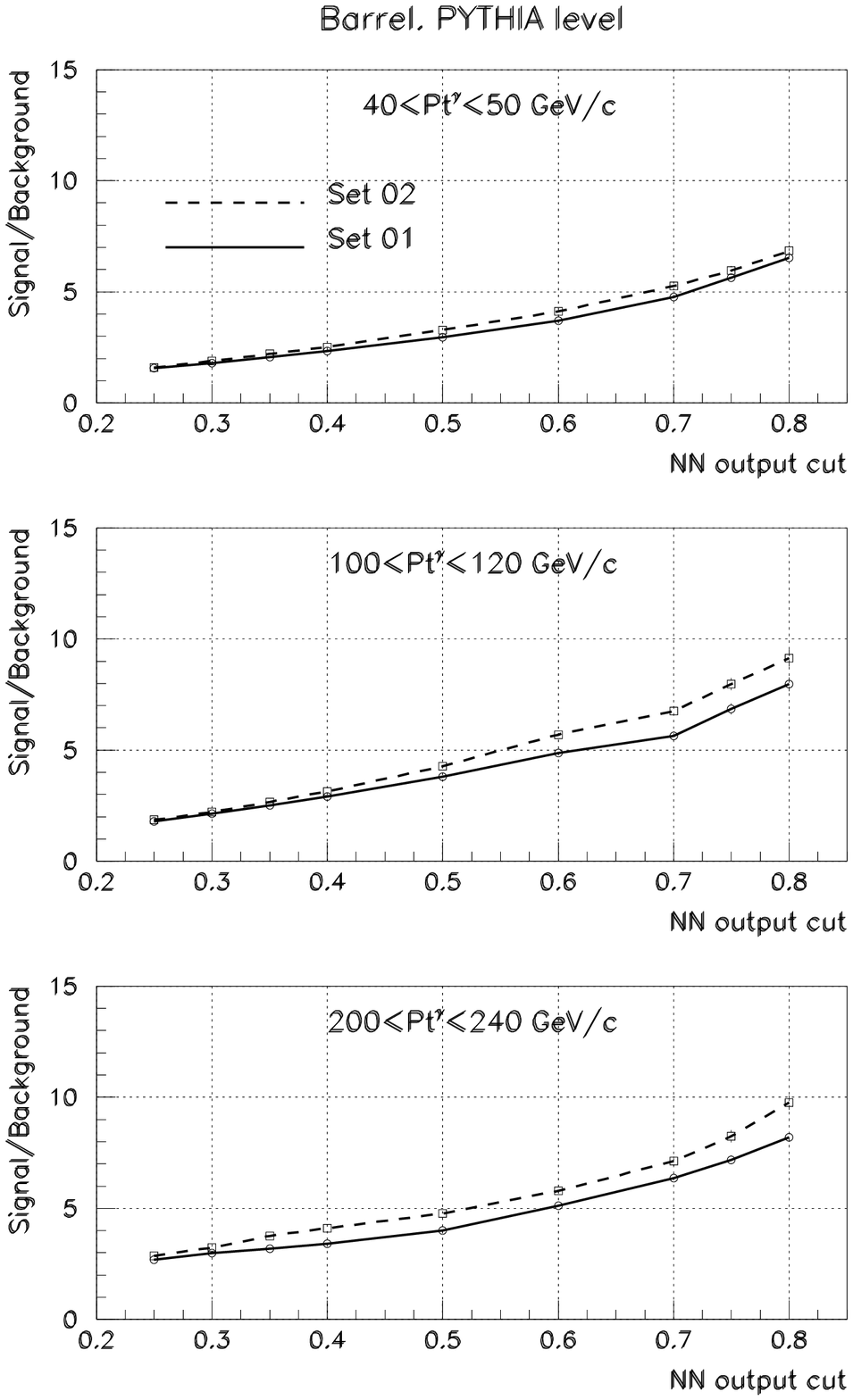}
    \vskip-.7cm
    \caption{\footnotesize Signal/Background ratio as a function of 
the NN output threshold value at the PYTHIA level.  Barrel region.}
    \label{fig:sb_hb} 
  \end{figure}
  \begin{figure}
    \includegraphics[width=1.0\textwidth,height=17.5cm]{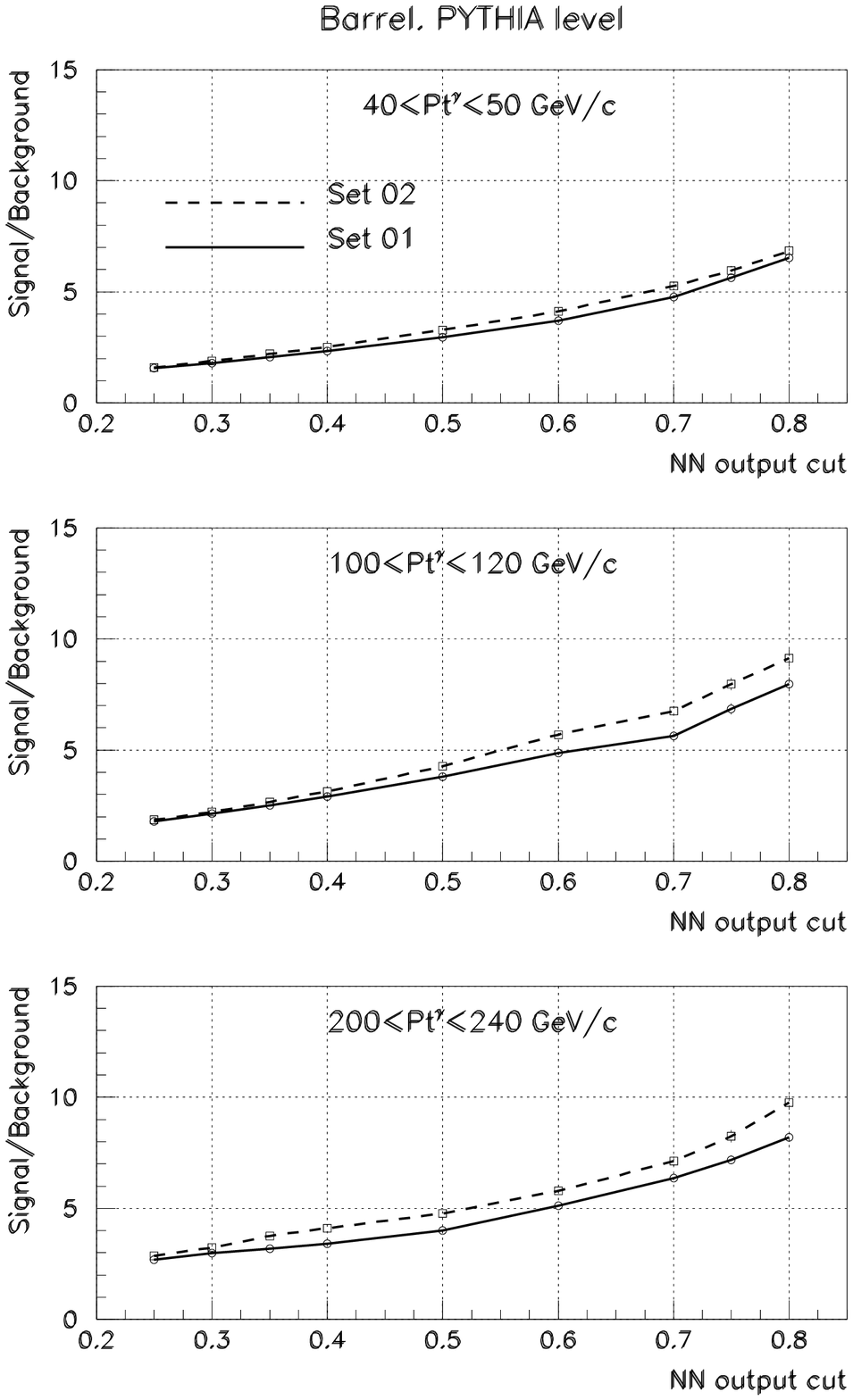}
    \vskip-.7cm
    \caption{\footnotesize  Signal/Background ratio as a function of 
the NN output threshold value at the PYTHIA level.  Endcap+Forward region.}
    \label{fig:sb_hef}
  \end{figure}

\end{document}